\documentclass[onecolumn]{emulateapj}

\usepackage{natbib}

\usepackage{color}
\definecolor{dr}{rgb}{0.8,0,0}
\definecolor{db}{rgb}{0,0,0.6}

\usepackage[ hyperfootnotes={false},
dvips,
pdftitle={Dispersal of protoplanetary disks by central wind stripping},
pdfauthor={Matsuyama, Johnstone, and Hollenbach},
pdfstartview={FitH},
bookmarksopen={true} ]{hyperref}
\hypersetup{colorlinks={true}, urlcolor={db}, citecolor={db}}

\begin{document}


\title{Dispersal of protoplanetary disks by central wind stripping}


\author{I. Matsuyama\altaffilmark{1}}
\affil{Department of Earth and Planetary Science, University of California, Berkeley, 307 McCone Hall, Berkeley, CA 94720, USA}
\email{isa@berkeley.edu}

\author{D. Johnstone\altaffilmark{2}}
\affil{National Research Council of Canada, Herzberg Institute of Astrophysics, 5071 West Saanich Road, Victoria, BC V9E 2E7, Canada}
\and

\author{D. Hollenbach\altaffilmark{3}}
\affil{SETI Institute, 515 N. Whisman Road, Mountain View, CA 94043, USA}


\altaffiltext{1}{Department of Terrestrial Magnetism, Carnegie Institution of Washington, 5241 Broad Branch Road NW, Washington, DC 20015}
\altaffiltext{2}{Department of Physics and Astronomy, University of Victoria, Victoria, BC V8P 1A1, Canada}
\altaffiltext{3}{NASA Ames Research Center,  Mail Stop 245-6, Moffett Field, CA 94035, USA}


\begin{abstract}
We present a model for the dispersal of protoplanetary disks by winds
from either the central star or the inner disk. These winds obliquely
strike the flaring disk surface and strip away disk material by entraining
it in an outward radial-moving flow at the wind-disk interface which
lies several disk scale heights above the mid-plane. The disk dispersal
time depends on the entrainment velocity, $v_{d}=\epsilon c_{s}$,
at which disk material flows into this turbulent shear layer interface,
where $\epsilon$ is a scale factor and $c_{s}$ is the local sound
speed in the disk surface just below the entrainment layer. If $\epsilon\sim0.1$,
a likely upper limit, the dispersal time at 1 AU is $\sim6$ Myr for
a disk with a surface density of $10^{3}$ g cm$^{-2}$, a solar mass
central star, and a wind with an outflow rate $\dot{M}_{w}=10^{-8}\mbox{ M}_{\odot}\mbox{ yr}^{-1}$
and terminal velocity $v_{w}=200\mbox{ km s}^{-1}$. When compared
to photoevaporation and viscous evolution, wind stripping can be a
dominant mechanism only for the combination of low accretion rates
($\lesssim10^{-8}$ M$_{\odot}$ yr$^{-1}$) and wind outflow rates
approaching these accretion rates. This case is unusual since generally
outflow rates are $\lesssim 0.1$ of accretion rates.
\end{abstract}





\section{Introduction}

A crucial timescale in planet formation is the timescale required
for the loss of protoplanetary gas material which initially dominates
the total disk mass. The outcome for a particular planetary system
might be very different if the parent disk is dispersed faster or
slower than in our solar system. However, relatively little attention
has been paid to the processes responsible for dispersing protoplanetary
disks. \citet{Hollenbach:1994} set a theoretical framework and \citet{Shu:1993}
applied it to propose photoevaporation of the solar nebula as the
disk dispersal mechanism capable of explaining the differences in
envelope masses between the gas-rich giants, Jupiter and Saturn, and
the gas-poor giants, Uranus and Neptune. \citet{Hollenbach:2000}
generalized the discussion, describing the variety of possible disk
dispersal mechanisms. The dominant disk dispersal mechanism at the
inner parts of the disk is viscous accretion onto the central star.
However, this process becomes inefficient with time as the outer disk
continuously expands to conserve angular momentum and the accretion
rate decreases. Other possible disk dispersal mechanisms are planet
formation, stellar encounters, stellar winds or disk winds, and photoevaporation
by energetic photons. Photoevaporation is a process in which the surface
of the disk is heated by stellar photons, resulting in a hydrodynamical
flow, a slow wind ($\lesssim10$ km s$^{-1}$), back to the interstellar
medium. \citet{Hollenbach:2000} concluded that planet formation
is a minor disk dispersal mechanism, and that the dominant mechanisms
for a wide range of disk sizes are viscous accretion in the inner
disk and photoevaporation in the outer disk. Therefore, planet formation
must compete with these more efficient dispersal mechanisms.

\citet{Handbury:1976} found that a stellar wind could not
have removed the solar nebula. They argue that the stellar wind could
only push the nebula to a finite distance determined by force balance
and angular momentum conservation. However, they assumed that the
stellar wind pushes the nebula as a whole and that the nebula maintains
Keplerian rotation. \citet{Yun:2007} studied the geometrical
and thermal structure of the wind-disk interface for the specific
case of a passive disk. However, they did not consider disk dispersal
by wind stripping. 
\citet{Cameron:1973} suggested that the interaction of an outflowing
stellar wind with the solar nebula would lead to significant mass loss, and 
\citet{Horedt:1978} and \citet{Elmegreen:1978} proposed models which
did consider disk dispersal. Although both of these models predict significant
mass loss, the manner in which the disk is dispersed is remarkably different. 
\citet{Horedt:1978} predicts that the wind drives an outward flow, while 
\citet{Elmegreen:1978} predicts, somewhat paradoxically, that the wind 
ultimately drives an inward flow. The difference between these two models is
discussed in \citet{Elmegreen:1979}. In particular, the wind-disk interface 
is given a priori in \citet{Horedt:1978}, while it is calculated in 
\citet{Elmegreen:1978} by considering normal pressure balance.
We follow the latter course here.
\citet{Elmegreen:1978} showed that if the momentum is deposited
where the wind strikes the disk, the addition of low angular momentum
wind material causes the underlying layers to spiral inwards. As
we will show below, if Keplerian rotation is assumed in the mixing
layer, the net radial flow is inward. In this case, the wind causes
disk dispersal by accelerating accretion onto the central star, as shown 
by \citet{Elmegreen:1978}.

We consider the alternative case of a shear mixing layer whose rotation
is non-Keplerian but conserves momentum of wind and entrained disk
surface material. The velocity shear is large and the tenuous surface
material moves outward at speeds greater than the gravitational escape
speeds. The disk material is expected to be entrained into the wind-disk
interface in this case, and is carried outwards to the interstellar
medium. The entrainment layer, or shear layer, carries with it a mixture
of shocked wind material as well as entrained disk material. 

The observationally inferred wind mass loss rate in T Tauri stars
can be as high as $\sim10^{-7}$ M$_{\odot}$ yr$^{-1}$ \citep[e.g. ][]{Cabrit:1990,Hartigan:1995,White:2004}.
For comparison, the present mass loss rate of the Sun's stellar wind
is $\sim10^{-14}$ M$_{\odot}$ yr$^{-1}$. The strong magnetic activity
in young stars is capable of driving stellar winds with outflow rates
$\lesssim10^{-8}$ M$_{\odot}$ yr$^{-1}$ \citep{Decampli:1981}.
In young stars with accretion disks, the interaction of the rotating
magnetic field with the accreting disk can generate outflows with
mass loss rates $\lesssim10^{-7}$ M$_{\odot}$ yr$^{-1}$ for correspondingly
high accretion rates $\lesssim10^{-6}$ M$_{\odot}$ yr$^{-1}$ \citep[see review by][]{Koenigl:1993}.
It remains unclear whether the outflows are launched near the magnetospheric
truncation radius \citep{Shu:2000} or over a wider range in disk
radii \citep{Konigl:2000}. Our emphasis here is on the subsequent
interaction of the wind, however launched, with the protoplanetary
disk. Thus, hereafter we will simply refer to the stellar wind or
the disk outflow as the central wind.

The rest of the paper is organized as follows. $\S$\ref{sec:theory}
describes the theory, which incorporates pressure balance; and mass,
momentum, and angular momentum conservation. $\S$\ref{sec:Results}
presents the results of applying the theory to protoplanetary disks.
Finally, $\S$\ref{sec:Discussion} summarizes the main results and
discusses some of their consequences.

\section{Model}\label{sec:theory}

\subsection{Wind-disk mixing layer}

\begin{figure}[!t]
\begin{centering}
\includegraphics[width=10cm]{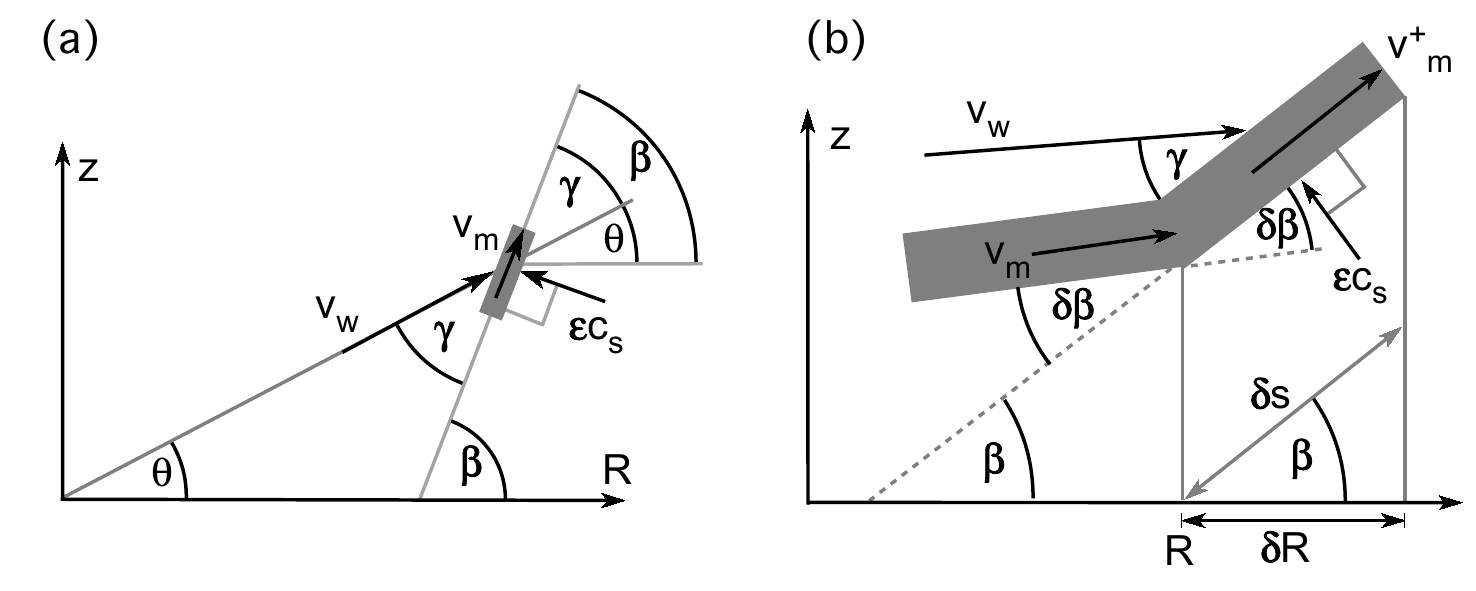}
\par\end{centering}
\caption{\label{cap:geometry}(a) Mass flow from the wind (with velocity $v_{w}$
and incidence angle $\gamma$) and the underlying disk (with velocity
$\epsilon c_{s}$ perpendicular to the mixing layer surface) to the
wind-disk mixing layer. There is mass transport through the mixing
layer in the direction of the local tangent angle, $\beta$, at velocity
$v_{m}$. (b) A mixing layer annulus of length $\delta s$ between
$R$ and $R+\delta R$ receives mass, momentum, and angular momentum
from the wind (with velocity $v_{w}$ and incidence angle $\gamma$),
the disk (with velocity $\epsilon c_{s}$ perpendicular to the mixing
layer surface), and the inner annulus (with velocity $v_{m}$ and
incidence angle $\delta\beta$), while it delivers mass and momentum
to the outer annulus (with velocity $v_{m}^{+}$ ). The local tangent
of the mixing layer surface is given by $dz/dR=\tan\beta$. }
\end{figure}

Following the analysis of \citet{Hollenbach:2000}, we assume that
disk dispersal by the stellar wind occurs in a thin mixing layer at
height $z(R)$, where $R$ is the disk radius. Disk material flows
into this mixing layer with speed $\epsilon c_{s}$, where $c_{s}$
is the sound speed and $\epsilon$ is the entrainment efficiency,
as shown in Figure \ref{cap:geometry}. This type of model has been
applied to protostellar jets entraining circumstellar material by
\citet{Canto:1991}. Since the disk cannot react to changes at velocities
faster than the sound speed, $\epsilon=1$ is a natural upper limit
to the possible value of the entrainment efficiency. Experimental
results at high Mach numbers in the regime that is relevant for central
winds suggest that $\epsilon\sim0.1$ is a more likely upper limit
\citep{Canto:1991}. We will consider entrainment efficiencies in
the range $0.01-1$. 

We denote the wind density, velocity, and incidence angle as $\rho_{w}$,
$v_{w}$, and $\gamma$ respectively; the disk density and Keplerian
angular velocity as $\rho_{d}$ and $\Omega_{k}$; and the mixing
layer surface density, velocity, inclination, and angular velocity
as $\Sigma_{m}$, $v_{m}$, $\beta$, and $\Omega_{m}$ respectively.
The wind incidence angle, $\gamma$, and the mixing layer inclination,
$\beta$, are related by \begin{equation}
\beta=\gamma+\theta,\label{eq:angles}\end{equation}
where $\tan\theta=z/R$ (Figure \ref{cap:geometry}). 

In Appendix \ref{sec:equations}, we derive equations for pressure
balance, and mass, momentum, and angular momentum conservation in
the mixing layer. For the benefit of the reader, we repeat these equations
here: 

\begin{eqnarray}
0 & = & \rho_{w}v_{w}^{2}\sin^{2}\gamma-\rho_{d}c_{s}^{2}(1+\epsilon^{2})+\Sigma_{m}v_{m}^{2}\cos\beta\frac{\partial\beta}{\partial R}\label{eq:pressure-balance}\\
\frac{\partial\Sigma_{m}}{\partial t} & = & \rho_{w}v_{w}\sin\gamma+\rho_{d}\epsilon c_{s}-\frac{\cos\beta}{R}\frac{\partial}{\partial R}\left(R\Sigma_{m}v_{m}\right)\label{eq:mass-cons}\\
\frac{\partial}{\partial t}\left(\Sigma_{m}v_{m}\right) & = & \rho_{w}v_{w}^{2}\sin\gamma\cos\gamma-\frac{\cos\beta}{R}\frac{\partial}{\partial R}\left(R\Sigma_{m}v_{m}^{2}\right)\label{eq:momentum-cons}\\
\frac{\partial}{\partial t}\left(\Sigma_{m}\Omega_{m}\right) & = & \rho_{d}\epsilon c_{s}\Omega_{k}-\frac{\cos\beta}{R^{3}}\frac{\partial}{\partial R}\left(R^{3}\Sigma_{m}v_{m}\Omega_{m}\right).\label{eq:ang-mom-cons}\end{eqnarray}

Equation (\ref{eq:pressure-balance}) describes normal pressure balance
in the mixing layer. On the right-hand-side (RHS) of this equation,
the first and second terms correspond to the normal pressures of the
wind and the disk respectively, while the third term describes the
centrifugal force associated with the mixing layer curvature. A similar,
so-called centrifugal correction, term has been considered in previous
studies (\citealp[p. 137, eq. 3.2.7]{Hayes:1966}; \citealp[eq. 10]{Canto:1980};
\citealp[eq. 7b]{Hartmann:1989}; \citealp[eq. 34]{Wilkin:1998}).
While these studies assume that the tangential momentum of the mixing
layer is given by the total accumulated tangential momentum of the
wind, we calculate it self-consistently by solving for $\Sigma_{m}$,
$v_{m}$, and $\beta$ from equations (\ref{eq:pressure-balance})-(\ref{eq:ang-mom-cons}).

\begin{figure}[!t]
\begin{centering}
\includegraphics[width=7cm]{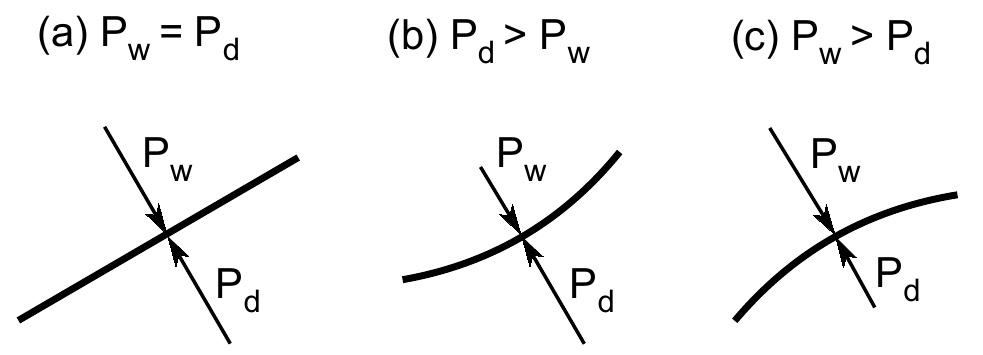}
\par\end{centering}
\caption{\label{cap:curva}Relationship between the curvature of the mixing
layer and normal pressure balance given by equation (\ref{eq:pressure-balance}).
The wind pressure is given by $P_{w}\equiv\rho_{w}v_{w}^{2}\sin^{2}\gamma$
and the total disk pressure is given by $P_{d}=\rho_{d}c_{s}^{2}(1+\epsilon^{2})$.}
\end{figure}

Figure \ref{cap:curva} illustrates the relationship between the mixing
layer curvature and normal pressure balance given by equation (\ref{eq:pressure-balance}).
If the wind pressure, $\rho_{w}v_{w}^{2}\sin^{2}\gamma$, is equal
to the total disk pressure, $\rho_{d}c_{s}^{2}(1+\epsilon^{2})$,
the curvature, $\cos\beta\partial\beta/\partial R$, is zero, as shown
in Figure \ref{cap:curva}a. If the total disk pressure is larger
than the wind pressure, the curvature is positive and the mixing layer
becomes concave on the side of the wind (i.e. the mixing layer surface
flares outward with increasing radius), as shown in Figure \ref{cap:curva}b.
Conversely, if the wind pressure is larger than the total disk pressure,
the curvature is negative and the mixing layer becomes concave on
the side of the disk, as shown in Figure \ref{cap:curva}c.

Equation (\ref{eq:mass-cons}) describes mass conservation. 
At a given radius, disk material mixes not only with the wind but also with the 
wind material already mixed with the disk gas at smaller radii.
On the RHS of this equation, the first and second terms describe mass input
from the wind and the disk respectively, and the third term describes
mass transport through the mixing layer. If we ignore the wind and
disk mass input, equation (\ref{eq:mass-cons}) is similar to the
standard continuity equation in cylindrical coordinates, \begin{equation}
\frac{\partial\Sigma_{m}}{\partial t}+\frac{1}{R}\frac{\partial}{\partial R}\left(\Sigma_{m}Rv_{m}\right)=0.\label{eq:mass-continuity}\end{equation}
The factor of $\cos\beta$ in our continuity equation (\ref{eq:mass-cons})
arises because of the non-zero, variable, inclination of the mixing
layer. Note that if we ignore the wind and disk contributions, equations
(\ref{eq:mass-cons}) and (\ref{eq:mass-continuity}) are equivalent
for $\beta=0$, as expected. 

Equation (\ref{eq:momentum-cons}) describes tangential momentum conservation.
The first and second terms on the RHS of this equation correspond
to the wind contribution and the tangential momentum transport through
the mixing layer respectively. Once again, the factor of $\cos\beta$
arises because of the non-zero, variable, inclination of the mixing
layer.

Equation (\ref{eq:ang-mom-cons}) describes angular momentum conservation.
The first term on the RHS of this equation corresponds to the angular
momentum contribution from the underlying disk, while the second term
describes the angular momentum flux through the mixing layer. We assume
that the central wind does not carry appreciable angular momentum.

Combining equations (\ref{eq:mass-cons}) and (\ref{eq:ang-mom-cons})
yields\begin{eqnarray}
v_{m}&=&\frac{R^{2}\left[\rho_{d}\epsilon c_{s}\left(\Omega_{k}-\Omega_{m}\right)-\Sigma_{m}\partial_{t}\Omega_{m}-\Omega_{m}\rho_{w}v_{w}\sin\gamma\right]}{\Sigma_{m}\cos\beta\partial_{R}\left(\Omega_{m}R^{2}\right)}.
\label{eq:vm-omega}\end{eqnarray}
If we assume Keplerian rotation for the mixing layer ($\Omega_{m}=\Omega_{k}$
and $\partial_{t}\Omega_{m}=0$), equation (\ref{eq:vm-omega}) shows
that the mass flow must be toward the central star ($v_{m}<0$), which
is essentially in agreement with the result of \citet{Elmegreen:1978}.
As we will show in the next section, outward flow is possible if we
relax the assumption of Keplerian rotation in the shearing mixing
layer at the interface of wind and disk. 

We can write the disk mass loss rate per unit time and disk gas surface
area as \begin{equation}
\dot{\Sigma}_{ws}=\frac{2\rho_{d}\epsilon c_{s}}{\cos\beta},\label{eq:dsdt}\end{equation}
where the factor of two accounts for mass loss from both sides of
the disk and the factor of $\cos\beta$ accounts for the inclination
of the mixing layer. The characteristic dispersal time at a given
radius due to mass loss to the mixing layer can be written as \begin{equation}
t_{ws}\equiv\frac{\Sigma_{d}}{\dot{\Sigma}_{ws}}=\frac{\Sigma_{d}\cos\beta}{2\rho_{d}\epsilon c_{s}}.\label{eq:tws}\end{equation}

\subsection{Disk}

We assume hydrostatic equilibrium in the vertical direction to write
the disk density as\begin{equation}
\rho_{d}(R,\, z)=\frac{\Sigma_{d}\exp\left[-z^{2}/(2H^{2})\right]}{\sqrt{2\pi}H\textrm{erf}[z_{m}/\sqrt{2}H]}\label{eq:rho-sigma-z}\end{equation}
for $z<z_{m}$, where $\Sigma_{d}$ is the gas surface density of
the disk, $z_{m}$ is the mixing layer height, $\mbox{erf}(x)$ is
the error function, and $H$ is the disk scale height. The disk scale
height is given by \begin{equation}
H(R)\equiv R\left(\frac{k_{B}T_{d}R}{GM_{*}\mu}\right)^{1/2},\label{eq:scale-H}\end{equation}
where $k_{B}$ is Boltzmann's constant, $T_{d}$ is the disk temperature
at radius $R$, $\mu=2.34\,m_{H}$ is the disk mean particle mass,
and $m_{H}$ is the mass of a  hydrogen atom.
For a thin flat disk, $T_d\propto R^{-3/4}$, while for a disk that flares outward (Eq. \ref{eq:scale-H}), 
a larger fraction of the stellar flux is intercepted and $T_d\propto R^{-1/2}$ \citep{Kenyon:1987}.
We assume a midplane disk temperature power law distribution, 
$T_{d}=100\mbox{ K}(R/AU)^{-1/2}$ \citep{DAlessio:1998}.

In the so-called {}``$\alpha$-disk'' theory \citep{Shakura:1973},
the viscosity, $\nu$, is parameterized as $\nu=\alpha c_{s}H,$ where
$\alpha$ is a dimensionless constant parameter.  
For our adopted disk temperature profile, viscous diffusion causes the surface density profile to  approach
$\Sigma_d\propto R^{-1} $, regardless of specific initial conditions \citep{Lynden-Bell:1974,Hartmann:1998}.
We adopt a power-law for the radial dependence of the disk gas surface density, \begin{equation}
\Sigma_{d}(R)=10^{3}\mbox{ g cm}^{-2}\left(\frac{R}{\mbox{AU}}\right)^{-1}.\label{eq:sigma-plaw}\end{equation}

The characteristic viscous evolution time
scale is given by \begin{equation}
t_{\nu}\sim\frac{R^{2}}{\nu}\sim0.4\mbox{ Myr}\left(\frac{R}{\mbox{AU}}\right)\left(\frac{\alpha}{10^{-3}}\right)^{-1}.\label{eq:tnu}\end{equation}
Comparison of predictions from similarity solutions with observed accretion rates and disk sizes 
suggests $\alpha\sim10^{-2}$ \citep{Hartmann:1998}, while the observed semimajor axis distribution of extrasolar planets suggests $\alpha\sim10^{-4}$ \citep{Ida:2005}. 
We will consider $\alpha$ values in this range.

The accretion rate can be written as 
\begin{eqnarray}
\dot{M}_{acc}&=&3\pi\nu\Sigma_{d}\sim3\times10^{-8}\mbox{ M}_{\odot}\mbox{ yr}^{-1}\left(\frac{\alpha}{10^{-2}}\right)\left(\frac{\Sigma_{0}}{10^{3}\mbox{ g cm}^{-2}}\right)\left(\frac{T_{0}}{100\mbox{ K}}\right)\left(\frac{M_{*}}{1\mbox{ M}_{\odot}}\right)^{-1/2},
\label{eq:accretion-rate}
\end{eqnarray}
where $\Sigma_{0}$ and $T_{0}$ are the surface density and disk
temperature at 1 AU. The radial velocity of the accretion flow is $\sim2$ cm s$^{-1}$$(\alpha/10^{-3})$.
Hence our assumption that only material and not radial momentum crosses
into the mixing layer. 

The inner boundary is chosen such that the wind velocity is equal
to the escape velocity from the star at this radius:\begin{eqnarray}
R_{min}&=&\frac{2GM_{*}}{v_{w}^{2}}\sim0.2\mbox{ AU}\left(\frac{M_{*}}{M_{\odot}}\right)\left(\frac{v_{w}}{100\mbox{ km s}^{-1}}\right)^{-2}.
\label{eq:Rmin}
\end{eqnarray}

\subsection{Central wind}

We consider outflow rates $\lesssim10^{-8}$ M$_{\odot}$ yr$^{-1}$
that can be driven by the stellar wind \citep{Decampli:1981}, as
well as higher outflow rates $\lesssim10^{-7}$ M$_{\odot}$ yr$^{-1}$
that can be driven by the interaction of the rotating magnetic field
with the accreting disk \citep{Shu:2000,Konigl:2000}. We do not consider
the initial acceleration of the wind since it is expected to strike
the disk surface after achieving terminal velocity. For simplicity,
we assume a spherically symmetric wind with an isotropic distribution
of density $\rho_{w}$ and velocity $v_{w}$. In this case, in a reference
frame centered on the star, the wind density can be written as 
\begin{eqnarray}
\rho_{w}&=&\frac{\dot{M}_{w}}{4\pi v_{w}r^{2}}\sim10^{-17}\mbox{ g cm}^{-3}\left(\frac{\dot{M}_{w}}{10^{-8}\mbox{ M}_{\odot}\mbox{ yr}^{-1}}\right)
\left(\frac{v_{w}}{200\mbox{ km}\mbox{ s}^{-1}}\right)^{-1}\left(\frac{r}{1\mbox{ AU}}\right)^{-2},
\label{eq:density-for-ballistic-wind}\end{eqnarray}
where $r^{2}\equiv R^{2}+z^{2}$ and $\dot{M}_{w}$ is the outflow
mass loss rate per unit time.

Our model could easily be extended to also include collimated winds.
We note that the assumption of spherical symmetry and isotropic distribution
is required only in the small solid angle subtended by the disk. 
If the wind is substantially collimated, then the true wind mass
loss rate is substantially higher than the $\dot{M}_{w}$ used in
our ``spherically symmetric'' models. 

\begin{figure}[!t]
\begin{centering}
\includegraphics[width=6cm]{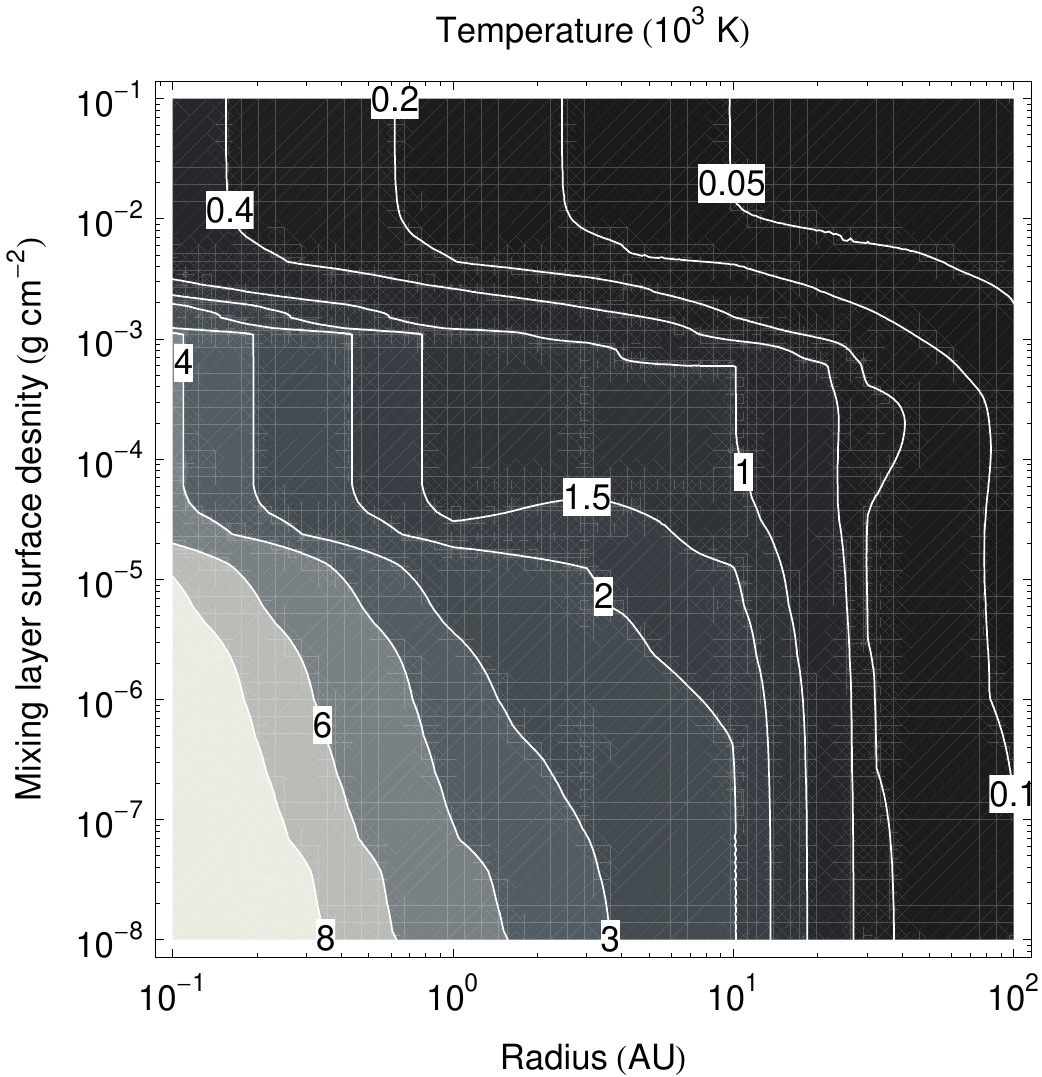}
\par\end{centering}
\caption{\label{fig:Temp}Contours of the gas temperature (in units of
$10^{3}$ K) in the mixing layer surface density-disk radius parameter
space. This is the surface temperature of the disk just below the
mixing layer.}
\end{figure}

\section{Results}\label{sec:Results}

We find steady state solutions (i.e. $\partial_{t}\Sigma_{m}=\partial_{t}\Sigma_{m}v_{m}=
\partial_{t}\Sigma_{m}\Omega_{m}=0$)
of equations (\ref{eq:mass-cons})-(\ref{eq:ang-mom-cons}) with the
boundary conditions $\Sigma_{m}(R_{min})=\Omega_{m}(R_{min})=0$ and
$\gamma(R_{min},z_{m}(R_{min}))=\gamma_{0}$, where $\gamma_{0}$
is a small initial wind incidence angle at $R_{min}$. We assume a
solar mass star, and a ballistic wind with $v_{w}=200\mbox{ km s}^{-1}$
and $\dot{M}_{w}=10^{-8}\mbox{ M}_{\odot}\mbox{ yr}^{-1}$ in our
fiducial model. 

We use the detailed thermo-chemical disk model of \citet{Gorti:2008}
to separately calculate the gas temperature, the sound speed, and
the gas density at the boundary between ambient disk and mixing layer.
This model includes FUV and X-ray radiation, dust collisions, photo-reactions,
and chemistry. For simplicity, we extrapolate calculations at specific
disk radii (1, 3, 10, 30, and 100 AU) to calculate the gas temperature
at different radii. Figure \ref{fig:Temp} shows contours of temperature
in the disk radius-surface density parameter space. Here, surface
density is measured from high $z$ downward toward the disk midplane.
The sharp transition at surface densities $\sim10^{-3}$ g cm$^{-2}$
is due to dust extinction of stellar photons, and the transition at
$\sim10^{-5}$ g cm$^{-2}$ is caused by a complicated interplay of
X-ray heating and ionization combined with [\ion{Ne}{2}] and {[\ion{Ar}{2}]
cooling. The transition at $\sim20$ AU is due to a relatively
sudden inability of FUV and X-ray heating to maintain gas temperatures
$\gg300$ K. Beyond this point {[\ion{O}{1}] 63 $\mu$m cooling can maintain
gas temperatures $\lesssim200$K. 

\begin{figure}[!t]
\begin{centering}
\includegraphics[width=15cm]{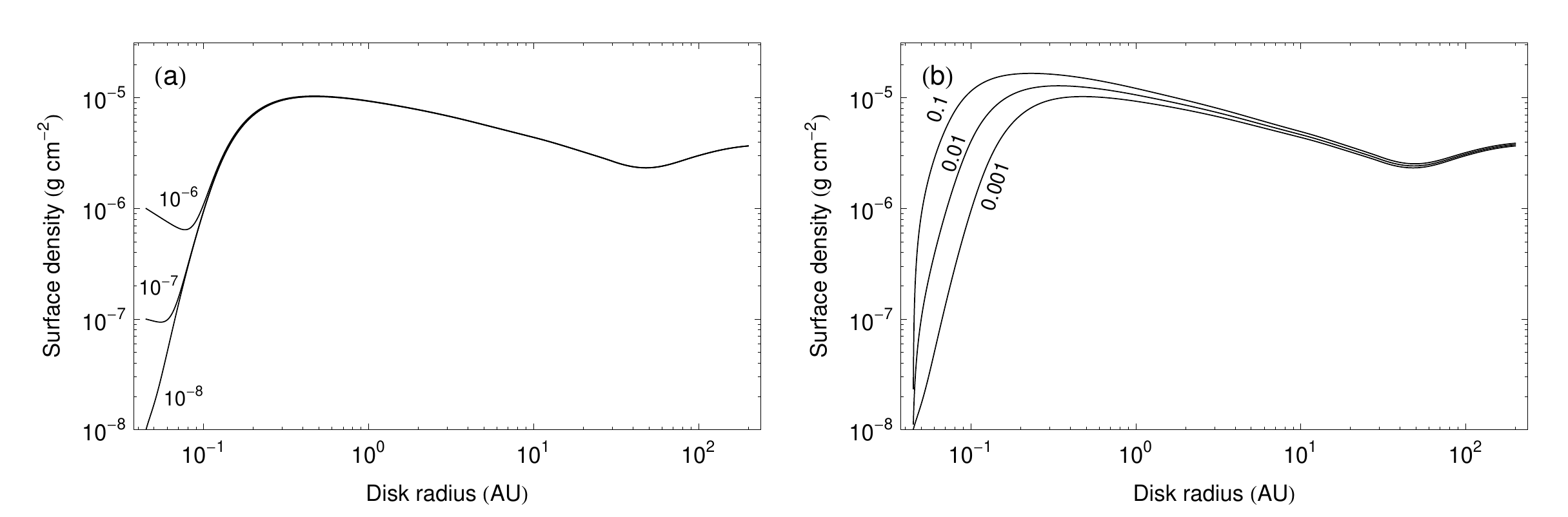}
\par\end{centering}
\caption{\label{fig:BCS}Mixing layer surface density as a function of disk
radius for different boundary conditions and an entrainment efficiency
$\epsilon=0.1$. (a) Solutions for a wind incidence angle of 0.001$^{\circ}$
and different mixing layer surface densities at the inner boundary
($10^{-6}$, $10^{-7}$, and $10^{-8}$ g cm$^{-2}$, as labeled on
each line). (b) Solutions for a mixing layer surface density of $10^{-8}$
g cm$^{-2}$ and different wind incidence angles at the inner boundary
(0.001, 0.01, and 0.1$^{\circ}$, as labeled on each line).}
\end{figure}

Figure \ref{fig:BCS} shows the mixing layer surface density for $\epsilon=0.1$,
a likely high efficiency, and different initial wind incidence angles
and mixing layer surface densities at the inner boundary varying over
several orders of magnitude. This figure illustrates that solutions
at large radii ($R\gtrsim$ 1 AU) are insensitive to the exact inner
boundary conditions used for a given value of $\epsilon$.

\begin{figure}[!tbph]
\begin{centering}
\includegraphics[width=17cm]{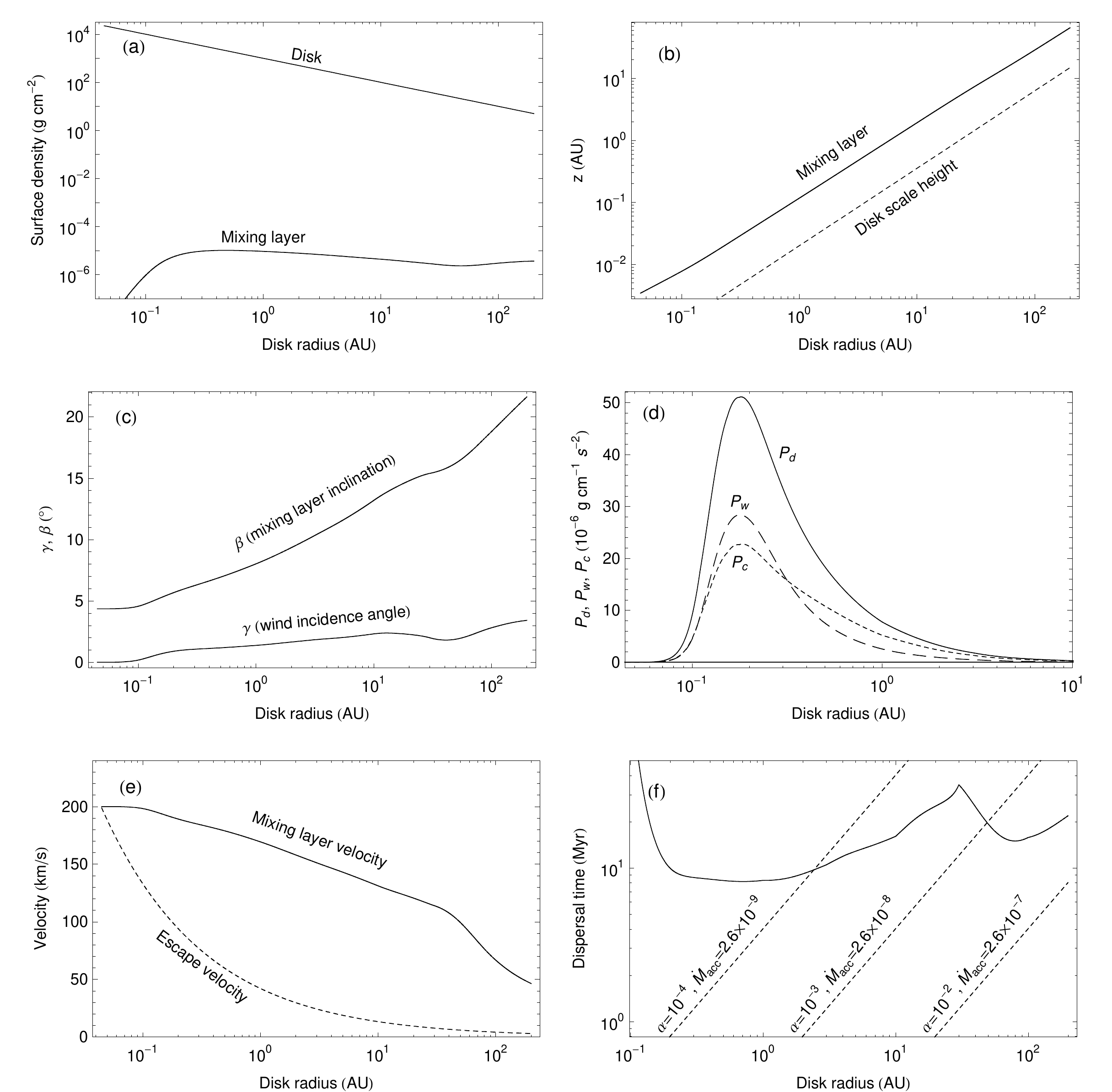}
\par\end{centering}
\caption{\label{fig:e01aall}Mixing layer and disk parameters as a function
of disk radius for an entrainment efficiency $\epsilon=0.1$. The
inner boundary conditions are a wind incidence angle of 0.001$^{\circ}$
and a zero mixing layer surface density. 
(a) Disk surface density, $\Sigma_{d}$, and mixing layer surface density, $\Sigma_{m}$, 
as labeled. 
(b) Mixing layer height (solid line) and disk scale height (dotted line). 
(c) Wind incidence angle, $\gamma$, and mixing layer inclination, $\beta$, as labeled. 
(d) Normal components of the disk pressure (solid line), 
$P_{d}\equiv\rho_{d}c_{s}^{2}(1+\epsilon^{2})$,
the wind pressure (dashed line), $P_{w}\equiv\rho_{w}v_{w}^{2}\sin^{2}\gamma$,
and the curvature pressure term (dotted line),
 $P_{c}\equiv\Sigma_{m}v_{m}^{2}\cos\beta d\beta/dR$. See equation (\ref{eq:pressure-balance}) 
 and Figure \ref{cap:curva} for a description of this pressure terms. 
(e) Mixing layer velocity (solid line), $v_{m}$, and escape velocity (dotted line). 
(f) Disk dispersal time (solid line), $\Sigma_{d}/\dot{\Sigma}_{ws}$, and viscous evolution
time scales (dotted lines) for different $\alpha$ viscosity values,
as labeled.}
\end{figure}

We summarize the main results for $\epsilon=0.1$ in Figure \ref{fig:e01aall}.
Although the wind-disk mixing layer accumulates material from the
wind and the underlying disk on the way out, surface density variations
in the mixing layer become small (at $R\gtrsim$ 1 AU, see Figure
\ref{fig:e01aall}a) due to the increasing surface area.

The wind-disk mixing layer is above the disk scale height (Figure
\ref{fig:e01aall}b) because the small wind incidence angles (Figure
\ref{fig:e01aall}c) result in correspondingly small wind ram pressures
that can be balanced by the small disk pressure at large height. The
disk pressure, $P_{d}\equiv\rho_{d}c_{s}^{2}(1+\epsilon^{2})$, is
larger than the normal wind pressure, $P_{w}\equiv\rho_{w}v_{w}^{2}\sin^{2}\gamma$,
(Figure \ref{fig:e01aall}d) and thus the mixing layer surface flares
outward with increasing radius (Figure \ref{fig:e01aall}c).

The mixing layer starts with roughly the same velocity as the wind
at the inner boundary (Figure \ref{fig:e01aall}e) since the wind
incidence angle is very small and very little disk mass has been entrained
(Figure \ref{fig:e01aall}c); and slows down at larger radii
as it accumulates mass from the disk (Figure \ref{fig:e01aall}e).
Although the mixing layer velocity decreases with radius, it remains
significantly higher than the escape velocity from the star. 

We compare the wind dispersal time (eq. {[}\ref{eq:tws}]) with the
viscous evolution time scale (eq. {[}\ref{eq:tnu}]) for a range of
accretion rates in Figure \ref{fig:e01aall}f. For our fiducial model
($\dot{M}_{w}=10^{-8}$ M$_{\odot}$ yr$^{-1}$, $v_{w}=200$ km s$^{-1}$,
and $\epsilon=0.1$) and an accretion rate $\sim3\times10^{-8}$ M$_{\odot}$
yr$^{-1}$, viscous evolution is the dominant mechanism in the inner
disk ($\lesssim50$ AU) and wind stripping is the dominant mechanism
in the outer disk in the absence of photoevaporation. We will explore
the conditions under which wind stripping dominates the disk evolution
in more detail below.

We derive an analytic expression for the dispersal time by assuming
that the curvature term contribution to the normal pressure balance
(eq. {[}\ref{eq:pressure-balance}]) is similar to that of the stellar
wind (Figure \ref{fig:e01aall}d). In this case, we can approximate
the normal disk pressure as $\rho_{d}c_{s}^{2}(1+\epsilon^{2})\sim2\rho_{w}v_{w}^{2}\sin^{2}\gamma$,
and the dispersal time (eq. {[}\ref{eq:tws}]) as 
\begin{equation}
t_{ws}\sim\frac{\pi\Sigma_{d}\cos\beta c_{s}R^{2}(1+\epsilon^{2})}{\dot{M}_{w}v_{w}\epsilon\sin^{2}\gamma},\label{tws:app1}\end{equation}
where we use equation (\ref{eq:density-for-ballistic-wind}) and $r\sim R$.
We replace $\beta\sim10^{\circ}$ and $\gamma\sim2^{\circ}$ (Figure
\ref{fig:e01aall}c) in equation (\ref{tws:app1}) to obtain
\begin{eqnarray}
t_{ws}(R)&\sim&6\mbox{ Myr}\left(\frac{\epsilon}{0.1}\right)^{-1}\left(1+\epsilon^{2}\right)\left(\frac{\Sigma_{0}}{10^{3}\mbox{ g cm}^{-2}}\right)\left(\frac{R}{\mbox{AU}}\right)
\times \left(\frac{c_{s}}{4\mbox{ km s}^{-1}}\right)\left(\frac{\dot{M}_{w}}{10^{-8}\mbox{ M}_{\odot}\mbox{yr}^{-1}}\right)^{-1}
\times \left(\frac{v_{w}}{200\mbox{ km s}^{-1}}\right)^{-1}.\label{eq:tws2}
\end{eqnarray}
Note that the sound speed decreases with radius and mixing layer surface
density (Figure \ref{fig:Temp}). The decreasing sound speed
means that $Rc_{s}$ does not increase much with $R$, which leads
to a fairly constant dispersal time as a function of $R$ in this
case where surface density drops with $R^{-1}$. Eq. {[}\ref{eq:tws2}]
is in agreement with the estimate of \citet{Hollenbach:2000}.
The highest dispersal rates (shortest dispersal times) correspond
to the highest wind outflow rates and velocities, as expected. We
can quantify the dispersing power of the wind with the parameter \begin{equation}
\eta\equiv\left(\frac{\epsilon}{0.1}\right)\left(\frac{\dot{M}_{w}}{10^{-8}\mbox{ M}_{\odot}\mbox{ yr}^{-1}}\right)\left(\frac{v_{w}}{200\mbox{ km s}^{-1}}\right),\label{eq:eta}\end{equation}
where we arbitrarily choose $\eta=1$ for our fiducial parameters. 

Figures \ref{fig:vs}-\ref{fig:param} compare results for models
with different entrainment efficiencies, wind outflow rates, and wind
velocities with our fiducial model ($\epsilon=0.1$, $\dot{M}_{w}=10^{-8}$
M$_{\odot}$ yr$^{-1}$, $v_{w}=200$ km s$^{-1}$). These results
agree well with the analytic approximation given in Eq. (\ref{eq:tws2}).

\begin{figure}[!t]
\begin{centering}
\includegraphics[width=14cm]{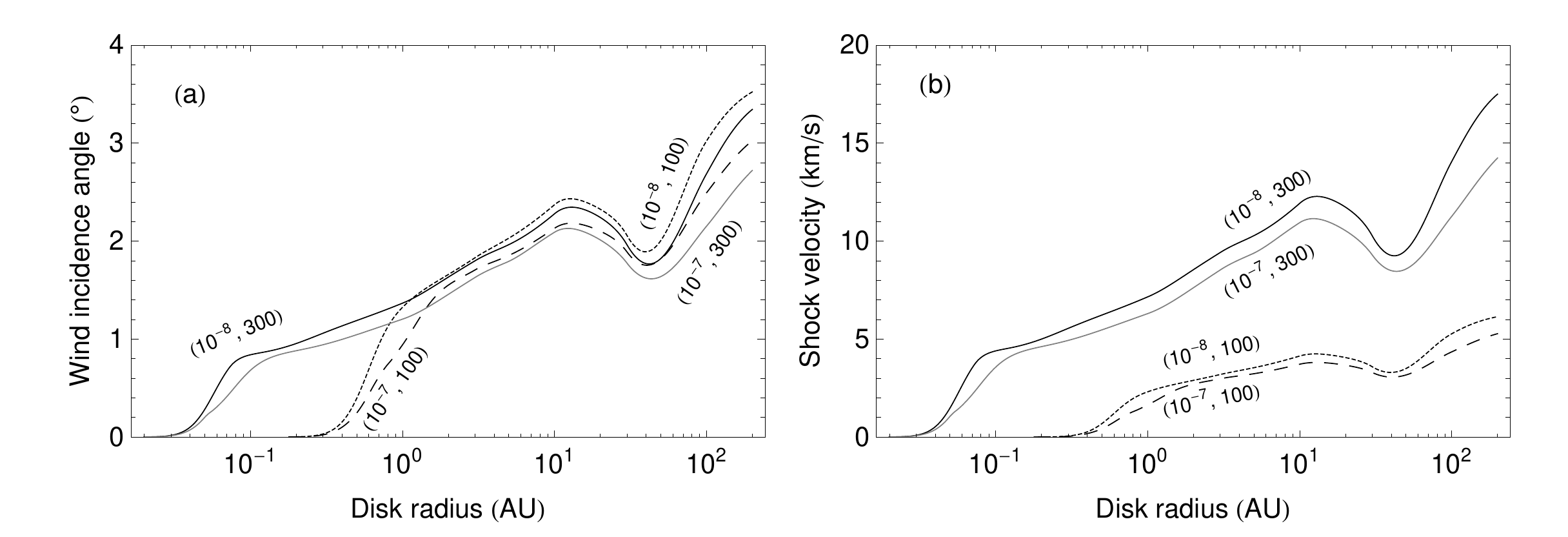}
\par\end{centering}
\caption{\label{fig:vs}Wind incidence angle (a), $\gamma$, and shock velocity
(b), $v_{s}=v_{w}\sin\gamma$, for an entrainment efficiency $\epsilon=0.1$,
stellar wind velocities in the range 100-300 km s$^{-1}$, and wind
outflow rates in the range $10^{-7}-10^{-8}$ M$_{\odot}$ yr$^{-1}$,
as labeled. Changing the entrainment efficiency results in negligible
variations.}
\end{figure}

We compare the wind incidence angles and the corresponding shock velocities
for $\epsilon=0.1$, wind outflow rates in the range $10^{-7}-10^{-8}$
M$_{\odot}$ yr$^{-1}$, and wind velocities in the range $100-300$
km s$^{-1}$ in Figure \ref{fig:vs}. Variations due to changes in
the entrainment efficiency are not significant. The wind incidence
angle is not sensitive to $\epsilon$, $\dot{M}_{w}$, or $v_{w}$
since the mixing layer is located several disk scale heights above
the midplane, where large changes in pressure (and mixing layer inclination)
can be accommodated by small changes in the mixing layer height. The
jogs at $\sim20$ AU are due to the rapid drop in gas temperature
beyond this radius (Figure \ref{fig:Temp}). We predict small wind
incidence angles $\gamma\lesssim4^{\circ}$ and correspondingly small
shock velocities (e.g. $v_{w}\sin\gamma=17$ km s$^{-1}$ at 100 AU
for $\dot{M}_{w}=10^{-8}$ M$_{\odot}$ yr$^{-1}$ and $v_{w}=300$
km s$^{-1}$). 

\begin{figure}[!t]
\begin{centering}
\includegraphics[width=14cm]{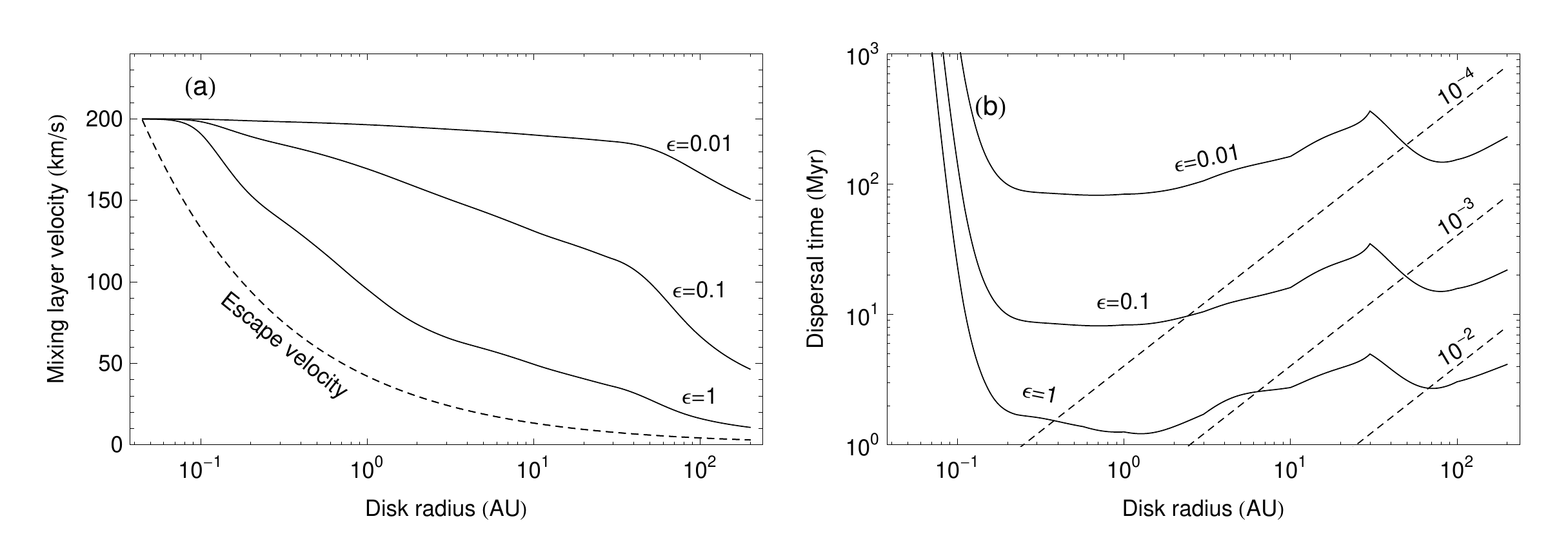}
\par\end{centering}
\caption{\label{fig:eall}Mixing layer velocity (a) and disk dispersal time
(b) as a function of disk radius for entrainment efficiencies in the
range $0.01<\epsilon<1$, as labeled. The other wind parameters are
standard ($\dot{M}_{w}=10^{-8}$ M$_{\odot}$ yr$^{-1}$ and $v_{w}=200$
km s$^{-1}$). The dotted lines in panel (b) give viscous evolution
time scales for different $\alpha$ viscosity parameters in the range
$10^{-4}<\alpha<10^{-2}$, as labeled.}
\end{figure}

We consider variations of the entrainment efficiency alone ($\epsilon=0.01-1$)
in Figure \ref{fig:eall}. Only the mixing layer velocity and the
dispersal time are shown since variations of the other quantities
shown in Figure \ref{fig:e01aall} are not significant. Increasing
the entrainment efficiency decreases the mixing layer velocity and
reduces the disk dispersal time, as expected since the mixing layer
receives more mass from the underlying disk in this case. Although
the mixing layer velocity decreases with radius as it entrains more
mass from the disk, it is possible to maintain the maximum possible
entrainment efficiency, $\epsilon=1$, since the mixing layer velocity
remains larger than the escape velocity at all radii.

Finally, we compare the dispersal mass flux rate and dispersal time
due to wind stripping with those due to photoevaporation and the mass
flux and characteristic time due to viscous evolution in Figure \ref{fig:param}.
Since the dispersal time depends on the particular disk surface density
distribution assumed, the mass flux outflow rate ({}``dispersal rate'')
is physically more meaningful. \citet{Hollenbach:2009} show that
EUV photons cannot penetrate the wind when the outflow rate is higher
than $\sim10^{-9}$ M$_{\odot}$ yr$^{-1}$, while FUV and X-ray photons
begin to penetrate once the outflow rate falls below $\sim10^{-7}$
M$_{\odot}$ yr$^{-1}$. Thus, we can ignore EUV photoevaporation
for the high outflow rates ($\gtrsim10^{-9}$ M$_{\odot}$ yr$^{-1}$)
considered here. We compare the dispersal rates by wind stripping
with those due to photoevaporation driven by FUV and X-ray heating
\citep[Figure 4]{Gorti:2009}. We assume outflow rate/accretion rate
ratios $\chi\equiv\dot{M}_{w}/\dot{M}_{acc}$ in the range $\sim0.01-1$,
as inferred from observations \citep[Figure 16, excluding edge-on disks]{White:2004}.
It is useful to consider different accretion rate regimes (Figure
\ref{fig:param}): 
\begin{itemize}
\item Low accretion rates ($\lesssim10^{-8}$ M$_{\odot}$ yr$^{-1}$, Figures
\ref{fig:param}a, b, c, and d). Viscous accretion is the dominant
mechanism at most disk radii ($\lesssim150$AU), while photoevaporation
is the dominant mechanism in the outermost regions ($\gtrsim150$
AU). Wind stripping can be dominant at $\sim100$ AU if the wind outflow
rate approaches the accretion rate ($\chi\sim1$). Once again, for
$\dot{M}_{w}\lesssim10^{-9}\mbox{ M}_{\odot}\mbox{ yr}^{-1}$, EUV
photons can penetrate the wind \citep{Hollenbach:2009} and EUV photoevaporation
becomes dominant at $\sim1$ AU \citep{Liffman:2003,Alexander:2006a}.

\item Intermediate accretion rates ($\sim10^{-7}$ M$_{\odot}$ yr$^{-1}$,
Figures \ref{fig:param}e and f). Viscous accretion is the dominant
mechanism in the inner disk ($R\lesssim60$ AU), while photoevaporation
is the dominant mechanism in the outer disk. Since the accretion rate also 
depends on the surface density (eq. [\ref{eq:accretion-rate}]), we assume a surface density 
higher than our fiducial value ($\Sigma_{d}=5\times10^{3}$ g cm$^{-2}$ at 1 AU) for 
intermediate and high accretion rates.

\item High accretion rates ($\gtrsim10^{-6}$ M$_{\odot}$ yr$^{-1}$, Figures
\ref{fig:param}g and h). Viscous accretion is the dominant mechanism
at all disk radii. In this case, although optical photons can reach and heat the disk, 
FUV and X-ray photons cannot penetrate the correspondingly high outflow rates 
\citep{Hollenbach:2009}.
\end{itemize}

\begin{figure}[!t]
\begin{centering}
\includegraphics[width=16cm]{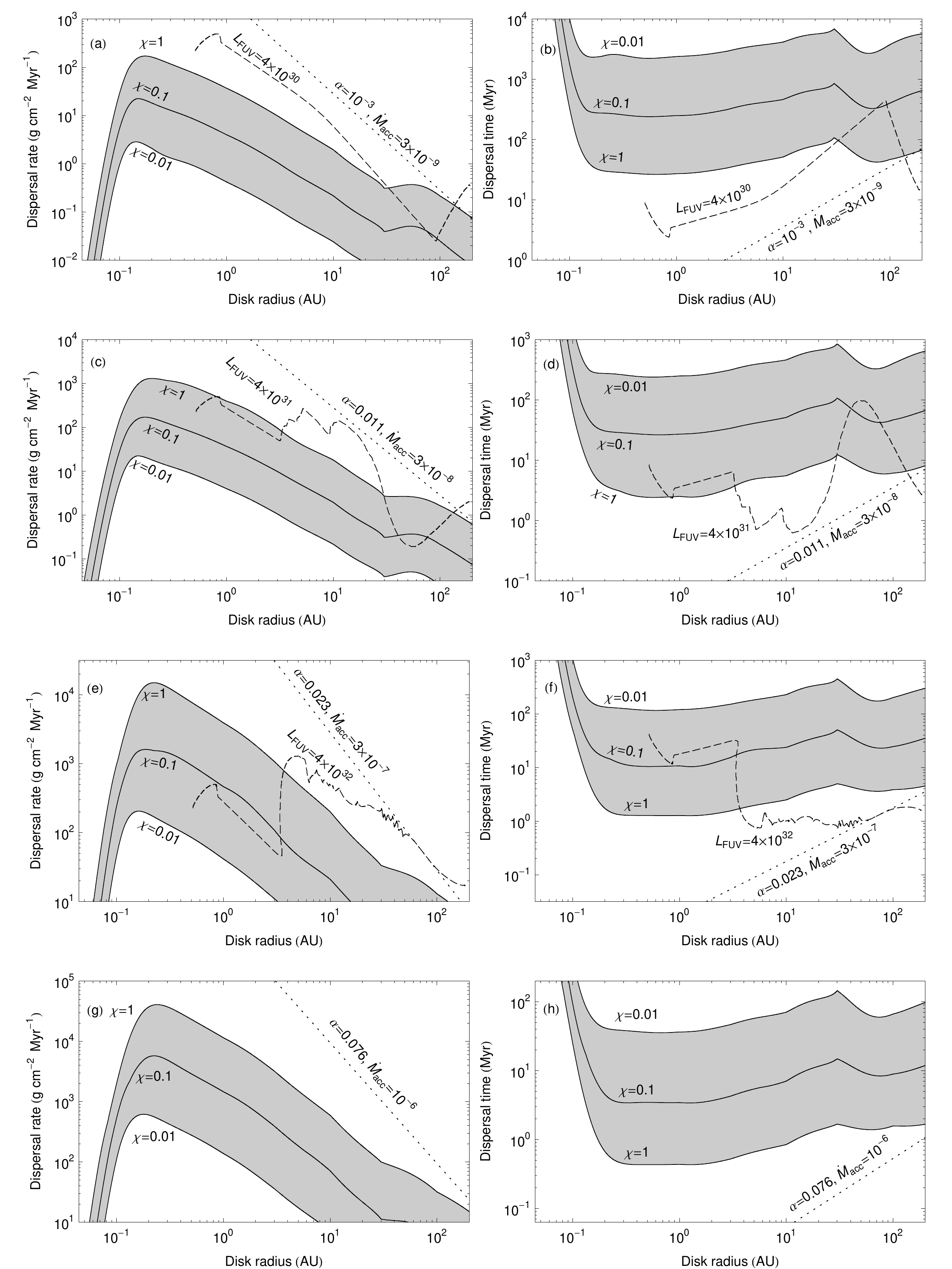}
\par\end{centering}
\caption{\label{fig:param}Dispersal rate and the corresponding dispersal time
(solid lines) for $\epsilon=0.1$, $v_{w}=200$ km s$^{-1}$, and
different outflow mass loss rate/accretion rate ratios $\chi\equiv\dot{M}_{w}/\dot{M}_{acc}$,
as labeled. Dotted lines show the mass transport rate and the corresponding
viscous evolution time scale due to accretion for accretion rates
in the range $3\times10^{-9}-10^{-6}$ M$_{\odot}$ yr$^{-1}$ (and
the corresponding $\alpha$ viscosity values using Eq. {[}\ref{eq:accretion-rate}]),
as labeled. Dashed lines show the photoevaporation dispersal rate
and the corresponding dispersal time (from Figure 4 of \citealp{Gorti:2009})
for the FUV luminosities, $L_{FUV}$, associated with the different
accretion rates \citep{Gorti:2008}, as labeled. We assume a surface
density higher than our fiducial value ($\Sigma_{d}=5\times10^{3}$
g cm$^{-2}$ at 1 AU) for correspondingly high accretion rates (e,
f, g, and h). }
\end{figure}

\section{Summary and conclusions}\label{sec:Discussion}

We present a model for the dispersal of protoplanetary disks by winds
from either the central star or the inner disk. These winds obliquely
strike the flaring disk surface and strip away disk material by entraining
it in an outward radial-moving flow at the wind-disk interface, located
at the surface of normal pressure balance several disk scale heights
above the mid-plane. We derive conservation equations for the mass,
momentum, and angular momentum in the wind-disk interface.

The disk dispersal time scale depends on the velocity at which
disk material is entrained into the mixing layer, which we quantify
with an entrainment velocity $\epsilon c_{s}$, where $c_{s}$ is
the sound speed and $\epsilon$ is an entrainment efficiency. The
dispersal time decreases as $(\epsilon v_{w}\dot{M}_{w})^{-1}$ (see
eq. {[}\ref{eq:tws2}]). For disk surface densities dropping as $R^{-1}$,
the dispersal time is relatively independent of $R$ (see Figs. \ref{fig:eall}
and \ref{fig:param} and Eq. {[}\ref{eq:tws2}]). We define a dimensionless
parameter $\eta\propto\epsilon v_{w}\dot{M}_{w}$ that is unity for
our standard parameters $\epsilon=0.1$, $v_{w}=200$ km s$^{-1}$,
and $\dot{M}_{w}=10^{-8}$ M$_{\odot}$ yr$^{-1}$. The dispersal
time for $\eta\sim1$ (eq. {[}\ref{eq:eta}]), $t_{ws}(1\mbox{ AU})\sim6(\Sigma_{0}/10^{3}\mbox{ g cm}^{-2})$
Myr, is somewhat larger than typical protoplanetary (dust) disk evolutionary
time scales ($\sim2$ Myr) inferred from infrared observations \citep[e.g.][]{Haisch:2001,Cieza:2007}.
In fact, as discussed below, viscous evolution and photoevaporation
likely dominate disk dispersal.

Figure \ref{fig:cartoon} depicts the dominant physical mechanisms
for different accretion rate regimes. For low accretion rates ($\lesssim10^{-8}$
M$_{\odot}$ yr$^{-1}$), viscous accretion is the dominant mechanism
at most disk radii ($\lesssim150$AU), while FUV and X-ray photoevaporation
is the dominant mechanism in the outermost regions ($\gtrsim150$
AU). In this case, wind stripping can be a dominant mechanism at $\sim100$
AU only if the wind outflow rates approach the accretion rates, which
is rare. EUV photoevaporation becomes a dominant mechanism
at $\sim1$ AU for low wind outflow rates ($\lesssim10^{-9}$ M$_{\odot}$
yr$^{-1}$). For intermediate accretion rates ($\sim10^{-7}$ M$_{\odot}$
yr$^{-1}$), viscous accretion is the dominant mechanism in the inner
disk ($\lesssim60$ AU), while FUV and X-ray photoevaporation is the
dominant mechanism in the outer disk. For high accretion rates ($\gtrsim10^{-6}$
M$_{\odot}$ yr$^{-1}$), viscous accretion dominates in the entire
disk. 

If wind stripping dominates viscous evolution, a gap may form if the
inward accretion flow is less than the mass flux produced by wind
entrainment at that radius, similar to the combined effects of viscous
accretion and EUV photoevaporation \citep{Clarke:2001,Matsuyama:2003a,Alexander:2006}.
If the wind mass loss rate scales with the accretion rate, such as
in X-wind models, then once a gap forms, accretion will rapidly decrease,
the wind will turn off, and the gap will be refilled by viscous evolution
of the outer disk. Time-dependent models which include both
the viscous evolution and mass loss by wind stripping are needed to
explore this scenario. Once again, this requires wind outflow rates
similar to the accretion rates (Figures \ref{fig:param}a, c). 

\citet{Hartmann:1989} found that shock heating at the wind-disk interface
can explain the observed forbidden line emission from large radii
($\sim50-100$ AU). However, they predict shock velocities $\sim20-30\mbox{ km s}^{-1}$
(see their Figure 6), while we predict smaller shock velocities ($\lesssim17$
km s$^{-1}$, Figure \ref{fig:vs}b) and correspondingly small wind
incidence angles ($\lesssim4^{\circ}$, Figure \ref{fig:vs}a), even
for wind velocities of $300$ km s$^{-1}$, higher than the maximum
value (200 km s$^{-1}$) assumed by \citet{Hartmann:1989}.
The difference arises because \citet{Hartmann:1989} assume that the
tangential momentum of the mixing layer is given by the total accumulated
tangential momentum of the wind while we solve for tangential momentum
conservation {[}eq. (\ref{eq:momentum-cons})]. Note that mass input
from the disk affects the tangential momentum of the mixing layer since it increases
the mixing layer surface density. The observed {[}\ion{Ne}{2}] 12.8 $\mu$m
forbidden line emission \citep[e.g.][]{Lahuis:2007} may be explained
by EUV or X-ray heating \citep{Glassgold:2007,Hollenbach:2009},
although the high {[}\ion{O}{1}] 6300 $\AA$ luminosities observed in some
cases \citep{Hartigan:1995} defy explanation. The combination of
EUV heating and ionization combined with turbulent mixing and heating
in the wind-induced shear layer may provide the high gas temperature
($\gtrsim3000$ K), high electron abundances, and high atomic O abundances
needed to explain the {[}\ion{O}{1}] luminosities. Further work is needed
in this area.

\begin{figure}[!t]
\begin{centering}
\includegraphics[width=10cm]{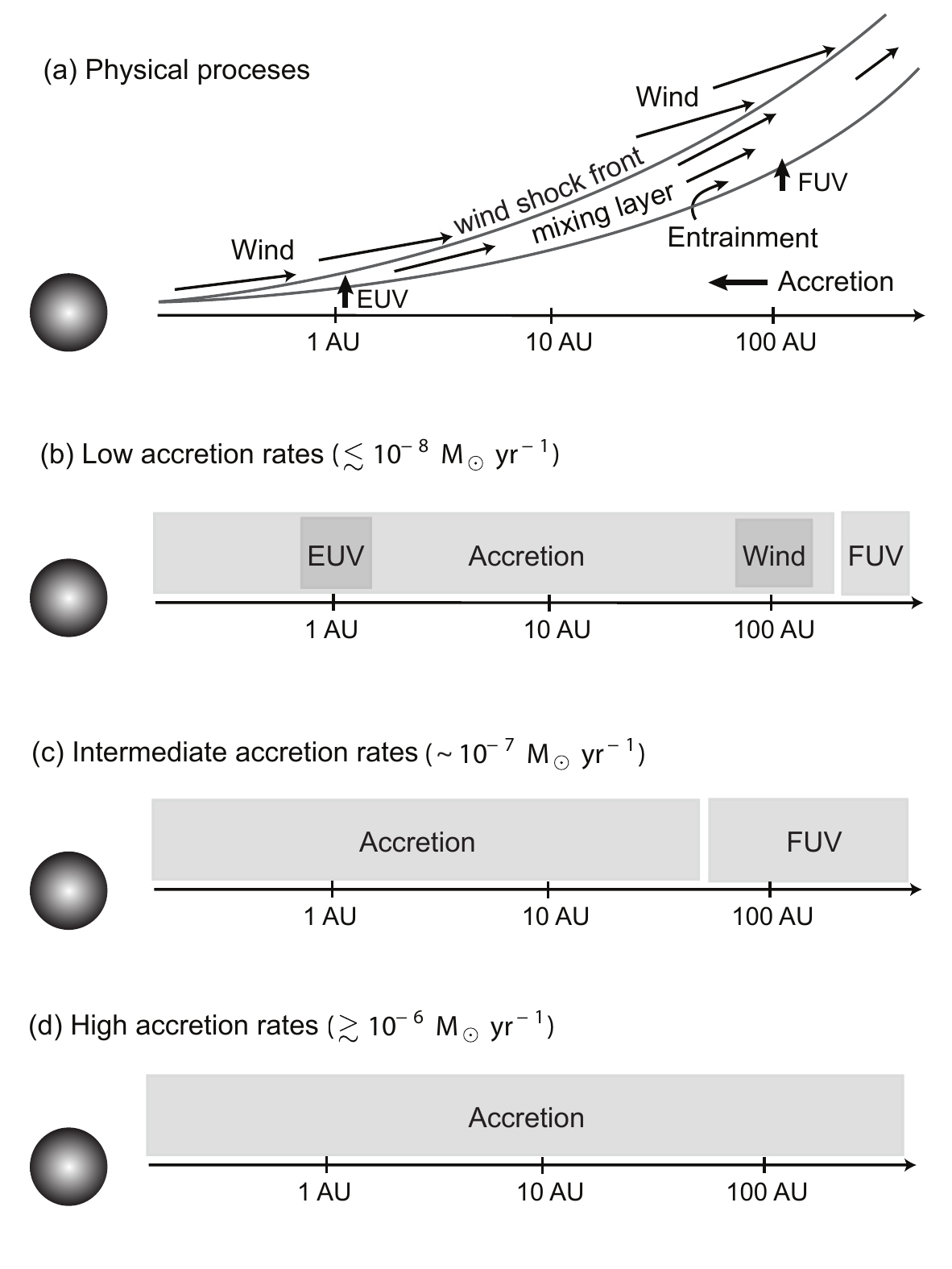}
\par\end{centering}
\caption{\label{fig:cartoon}Schematic illustrations of the physical processes
that dominate disk dispersal (a); and the accretion rate and disk
radii regimes in which viscous accretion, EUV photoevaporation
(EUV), FUV and X-ray photoevaporation (FUV), and wind stripping (wind)
are dominant mechanisms (b, c, d). 
The central wind interacts directly with the disk at all radii since the mixing layer flares outward (Fig. \ref{fig:e01aall}c).
In the mixing layer at a given radius, disk material mixes not only with the wind but also with the wind material already mixed with the disk gas at smaller radii.
Note that in (b), EUV dominates at $\sim1$
AU only when the wind mass loss rate $\lesssim10^{-9}$ M$_{\odot}$
yr$^{-1}$. }
\end{figure}

\acknowledgments
We thank Uma Gorti for supplying us with prepublication results of her disk models that allowed 
us to estimate the temperature structure of the disk just below the shear layer created by the wind. This research was supported by the  Miller Institute for Basic Research in Science, University of California, Berkeley; the Carnegie Institution of Washington; a Natural Sciences and Engineering Research Council of Canada grant; the NASA Astrophysical Theory program; and the NASA Astrobiology Institute.

\clearpage





%

\appendix



\section{Conservation equations}\label{sec:equations}

We consider normal pressure balance; and mass, momentum, and angular
momentum conservation for a mixing layer annulus between $R$ and
$R+\delta R$, as shown in Figure \ref{cap:geometry}. We denote the
wind density, velocity, and incidence angle as $\rho_{w}$, $v_{w}$,
and $\gamma$ respectively; the disk density and sound speed as $\rho_{d}$
and $c_{s}$ respectively, and the mixing layer surface density, velocity,
and inclination as $\Sigma_{m}$, $v_{m}$, and $\beta$ respectively. 

Force balance in the direction perpendicular to the surface of the
annulus can be written as

\begin{equation}
2\pi R\delta s\rho_{w}v_{w}^{2}\sin^{2}\gamma+2\pi R\Sigma_{m}v_{m}^{2}\sin\delta\beta=2\pi R\delta s\rho_{d}c_{s}^{2}+2\pi R\delta s\rho_{d}\epsilon^{2}c_{s}^{2}.\label{app:press1}\end{equation}
On the left-hand-side (LHS) of equation (\ref{app:press1}), the
first and second terms correspond to the force associated with the
normal momentum flux of the wind and the normal momentum flux through
the inner boundary respectively (Figure \ref{cap:geometry}). On the
RHS of equation (\ref{app:press1}), the first and second terms correspond
to the thermal and ram pressure forces of the disk respectively. Replacing
$\delta s=\delta R/\cos\beta$ and taking the limits $\delta R\rightarrow0$
and $\delta\beta\rightarrow0$ in equation (\ref{app:press1}) yields
\begin{equation}
\rho_{w}v_{w}^{2}\sin^{2}\gamma+\Sigma_{m}v_{m}^{2}\cos\beta\frac{\partial\beta}{\partial R}=\rho_{d}c_{s}^{2}(1+\epsilon^{2}).\label{app:press}\end{equation}
The second term on the LHS of equation (\ref{app:press}) is given
by the product of the tangential momentum flux per unit length and
time, $\Sigma_{m}v_{m}^{2}$, and the curvature of the mixing layer
surface, $\cos\beta d\beta/dR=d\beta/ds$. 

The mass and momentum in the direction tangent to the surface are
$2\pi R\delta s\Sigma_{m}$ and $2\pi R\delta s\Sigma_{m}v_{m}$ respectively.
The mass change per unit time is given by mass input from the wind
and the disk, and mass transport through the annulus: \begin{eqnarray}
\frac{\partial}{\partial t}\left(2\pi R\Sigma_{m}\delta s\right) & = & 2\pi R\delta s\rho_{w}v_{w}\sin\gamma+2\pi R\delta s\rho_{d}\epsilon c_{s}+2\pi R\Sigma_{m}v_{m}-2\pi R^{+}\Sigma_{m}^{+}v_{m}^{+},\label{app:mass1}\end{eqnarray}
where we define $R^{+}\equiv R+\delta R$, $\Sigma_{m}^{+}=\Sigma_{m}(R^{+},\, z(R^{+}),\, t)$,
and $v_{m}^{+}=v_{m}(R^{+},\, z(R^{+}),\, t)$. The last two terms
on the RHS of equation (\ref{app:mass1}) correspond to the mass flux
per unit time across the inner and outer boundaries of the annulus
respectively. Replacing $\delta s=\delta R/\cos\beta$ (Figure \ref{cap:geometry})
and taking the limit $\delta R\rightarrow0$ in equation (\ref{app:mass1})
yields \begin{equation}
\frac{\partial\Sigma_{m}}{\partial t}=\rho_{w}v_{w}\sin\gamma+\rho_{d}\epsilon c_{s}-\frac{\cos\beta}{R}\frac{\partial}{\partial R}\left(R\Sigma_{m}v_{m}\right).\label{app:mass}\end{equation}

Momentum conservation can be written as \begin{eqnarray}
\frac{\partial}{\partial t}\left(2\pi R\Sigma_{m}\delta sv_{m}\right) & = & 2\pi R\delta s\rho_{w}v_{w}^{2}\sin\gamma\cos\gamma+2\pi R\Sigma_{m}v_{m}^{2}\cos\delta\beta-2\pi R^{+}\Sigma_{m}^{+}v_{m}^{+2},\label{app:mom1}\end{eqnarray}
where the last two terms on the RHS are the momentum flux per unit
time across the inner and outer boundary of the annulus respectively.
Replacing $\delta s=\delta R/\cos\beta$ and taking the limits $\delta\beta\rightarrow0$
and $\delta R\rightarrow0$ in equation (\ref{app:mom1}) yields\begin{equation}
\frac{\partial}{\partial t}\left(\Sigma_{m}v_{m}\right)=\rho_{w}v_{w}^{2}\sin\gamma\cos\gamma-\frac{\cos\beta}{R}\frac{\partial}{\partial R}\left(R\Sigma_{m}v_{m}^{2}\right).\label{app:momentum}\end{equation}

Assuming that the wind does not carry significant angular momentum,
angular momentum conservation can be written as

\begin{eqnarray}
\frac{\partial}{\partial t}\left(2\pi R\Sigma_{m}\delta s\Omega_{m}R^{2}\right) & = & 2\pi R\delta s\rho_{d}\epsilon c_{s}\Omega_{k}R^{2}+2\pi R\Sigma_{m}v_{m}\Omega_{m}R^{2}-2\pi R^{+}\Sigma_{m}^{+}v_{m}^{+}\Omega_{m}^{+}R^{+2},
\label{app:ang-mom1}\end{eqnarray}
where $\Omega_{k}=(GM/R^{3})^{1/2}$ is the Keplerian angular velocity
of the disk and $\Omega_{m}$ is the angular velocity of the mixing
layer. The angular velocity of the disk must be sub-Keplerian due
to radial pressure support, by a factor $\sim1-\mathcal{O}(H/R)^{2}$,
where $H$ is the disk scale height. We ignore this effect since it
is negligible for protoplanetary disks with $(H/R)^{2}\ll1$. The
angular momentum loss due to the outward motion of the mixing layer
{[}third and fourth terms on the RHS of equation (\ref{app:ang-mom1})]
is partially compensated by the angular momentum input from the underlying
disk {[}first term on the RHS of equation \ref{app:ang-mom1}]. Once again, replacing
$\delta s=\delta R/\cos\beta$ and taking the limits $\delta\beta\rightarrow0$
and $\delta R\rightarrow0$ in equation (\ref{app:ang-mom1}) yields
\begin{equation}
\frac{\partial}{\partial t}\left(\Sigma_{m}\Omega_{m}\right)=\rho_{d}\epsilon c_{s}\Omega-\frac{\cos\beta}{R^{3}}\frac{\partial}{\partial R}\left(R^{3}\Sigma_{m}v_{m}\Omega_{m}\right).
\label{app:ang-mom}
\end{equation}





\clearpage

\end{document}